\documentclass[12pt]{article}
\usepackage{amssymb,amsmath}
\usepackage{caption}
\usepackage[noblocks]{authblk}
\begin{document}
\textwidth 10.0in 
\textheight 9.0in 
\topmargin -0.60in

\title{ The Renormalization Group and the Effective Action}
\author{D.G.C. McKeon}
\affil {Department of Applied Mathematics, The
University of Western Ontario, London, ON N6A 5B7, Canada} 
\affil {Department of Mathematics and
Computer Science, Algoma University, Sault St.Marie, ON P6A
2G4, Canada}

\maketitle

\maketitle 
\let\oldthefootnote\thefootnote
\renewcommand{\thefootnote}{\fnsymbol{footnote}}
\footnotetext[1]{Email: dgmckeo2@uwo.ca}
\let\thefootnote\oldthefootnote

\begin{abstract}
The renormalization group is used to sum the leading-log (LL) contributions to the effective action for a large constant external gauge field 
in terms of the one-loop renormalization group (RG) function $\beta$, the next-to-leading-log (NLL) contributions in terms of the two-loop RG 
function etc.  The log independent pieces are not determined by the RG equation, but can be fixed by considering the anomaly in the trace of the
energy-momentum tensor. Similar considerations can be applied to the effective potential $V$ for a scalar field $\phi$; here the log independent 
pieces are fixed by the condition $V^\prime (\phi = v) = 0$.
\end{abstract}

The effective Lagrangian for a constant external gauge field has been considered in a number of papers [1-7].  In the limit of a strong external 
field strength, these lead to logarithmic corrections to the classical Lagrangian $-\frac{1}{4} F_{\mu\nu}^2 \equiv -\frac{1}{4} \Phi$. (We can regard $F$ 
as being either the electromagnetic field strength in QED or a non-Abelian field strength which may be coupled to matter.)  A systematic summation 
of these effects by using the RG equation has been discussed in ref. [8]; the summation of logarithmic effects arising due to radiative processes 
in other contents has been considered in [9].  In this note we show that the RG equation, when applied to the effective Lagrangian $L$ for a strong 
external gauge field, can be rewritten as a sequence of coupled ordinary differential equations for functions $S_n$ with $S_0$ giving the LL 
contribution to $L$, $S_1$ the NLL contribution to $L$, etc.  The boundary conditions for these equations are the log-independent contributions to $L$.  
These can be fixed by examining the anomaly in the energy momentum tensor [10] as this anomaly can be used to find a formal expression for $L$ [11].  The 
approach used is similar to one employed with the effective potential when there is a fundamental scalar field in the model [12].

If $\mu$ is the renormalization scale in a model, $F_{\mu\nu}$ is the constant external field strength and $\lambda$ the gauge coupling, then the effective 
Lagrangian $L(F_{\mu\nu}, \lambda , \mu)$ must be independent of $\mu$ and hence the RG equation follows,
\begin{equation}
\frac{dL}{d\mu} = \left( \mu \frac{\partial}{\partial \mu} + \beta (\lambda) \frac{\partial}{\partial\lambda} + \gamma(\lambda)F_{\mu\nu} \frac{\partial}
{\partial F_{\mu\nu}}\right) L = 0.
\end{equation}
Since $\lambda F_{\mu\nu}$ is not renormalized, $\beta(\lambda) = - \lambda\gamma(\lambda)$ [13] and so if $\Phi = F^{\mu\nu}F_{\mu\nu}$, eq. (1) becomes 
\begin{equation}
\left[ \mu \frac{\partial}{\partial\mu} + \beta(\lambda)\left(\frac{\partial}{\partial\lambda} - \frac{2}{\lambda} \Phi \frac{\partial}{\partial\Phi}\right)\right]
L = 0.
\end{equation}
With $t = \frac{1}{4} \ln \left(\frac{\lambda^2\Phi}{\mu^4}\right)$, the form of $L$ when $\lambda\Phi >> \mu^2$ is [8]
\begin{equation}
L = \sum_{n=0}^\infty \sum_{m=0}^n \,T_{n,m} \lambda^{2n}t^m\Phi = \sum_{n=0}^\infty S_n(\lambda^2t)\lambda^{2n}\Phi
\end{equation}
where $S_n(\lambda^2t) = \displaystyle{\sum_{m=0}^\infty} T_{n+m,m}(\lambda^2t)^m$.  If $\beta(\lambda) = \displaystyle{\sum_{n=1}^\infty} b_{2n+1}\lambda^{2n+1}$, eq. (2) is satisfied 
at progressively higher orders in $\lambda$ provided these functions satisfy a set of coupled ordinary differential equations,
\begin{subequations}
    \begin{eqnarray}
wS_0^\prime(w)\!\!&-&\!\!S_0 = 0\\
b_3wS_1^\prime(w)\!\! &-&\!\! b_5S_0(w) + (1+w)b_5S_0^\prime(w) = 0\\
-b_3S_2^\prime +b_3S_2\!\! &-&\! \!b_7S_0 + (1+w)(b_7S_0^\prime +b_5S_1^\prime + b_3S_2^\prime) = 0
\end{eqnarray}
\end{subequations}
etc. with $w = -1 + 2b_3(\lambda^2t)$.  In general $S_n(\xi)$ can be found once $S_0 \ldots S_{n-1}$ have been determined provided $b_3 \ldots b_{2n+3}$ are known 
and the boundary conditions $S_n(\lambda^2t = 0) = T_{n0}$ have been specified.  In particular, $S_0 = T_{00}w$, $S_1 = -\frac{T_{00}b_5}{b_3} \ln|w| + T_{10}$ and 
$S_2 = \left[\left(\frac{b_5}{b_3}\right)^2 - \frac{b_7}{b_3}\right]T_{00}\ln|w| - \left(\frac{b_3}{b_5}\right)^2 T_{00}\left(\frac{1}{w} - 1\right) + T_{20}$.

To find these boundary conditions, an extra condition must be found.  To do this, we reexpress $L$ in eq. (3) as 
$L = \displaystyle{\sum_{n=0}^\infty} A_n(\lambda)t^n\Phi$ where\linebreak $A_n = \displaystyle{\sum_{m=n}^\infty} T_{m,n}\lambda^{2m}$.  Eq. (2) is now 
satisfied at each order in $t$ provided 
\begin{equation}
\frac{1}{\lambda^2} A_{n+1}(\lambda) = \frac{1}{n+1}\beta(\lambda)\frac{d}{d\lambda}\left(\frac{1}{\lambda^2} A_n(\lambda)\right).
\end{equation}
If now $A_n(\lambda) = \lambda^2\overline{A}_n(\lambda)$ and $\eta = \displaystyle{\int_{\lambda_0}^{\lambda(\eta)}} \frac{dx}{\beta(x)}$ then
\begin{equation}
\overline{A}_{n+1}(\lambda(\eta)) = \frac{1}{(n+1)!}\, \frac{d^{n+1}}{d\eta^{n+1}} \overline{A}_0(\lambda(\eta))
\end{equation}
so that
\begin{equation}
L = \lambda^2(\eta) \sum_{n=0}^\infty \frac{t^n}{n!}\,\frac{d^n}{d\eta^n} \overline{A}_0(\lambda(\eta))\Phi = \lambda^2(\eta)\overline{A}_0(\lambda(\eta + t))\Phi = 
\frac{\lambda^2(\eta)}{\lambda^2(\eta + t)} A_0(\lambda(\eta + t))\Phi\,.
\end{equation}
Since $A_0$ is determined by the $T_{n0}$, we see from eq. (7) that again the log independent contributions to $L$ fix the log dependent contributions once 
$\beta$ is known. When $\eta = 0$, we take the value of the function $\lambda(\eta)$ to be $\lambda_0$.

We now recall that the trace anomaly of the energy momentum tensor [10]
\begin{equation}
\left\langle \theta^\mu_{\,\,\mu}\right\rangle = \frac{\beta(\overline{\lambda})}{2\overline{\lambda}(t)}\,\frac{\lambda^2_0}{\overline{\lambda}^2(t)}\Phi
\end{equation}
where
\begin{equation}
\frac{d\overline{\lambda}(t)}{dt} = \beta (\overline{\lambda}(t))\;\;\left(t = \int_{\lambda_0}^{\overline{\lambda}(t)} \,\frac{dx}{\beta(x)}\right)
\end{equation}
leads to [11]
\begin{equation}
L = - \frac{1}{4} \frac{\lambda^2_0}{\overline{\lambda}^2(t)} \Phi
\end{equation}
since $\left\langle \theta^{\mu\nu}\right\rangle = -\eta^{\mu\nu} L+ 2 \frac{\partial L}{\partial\eta_{\mu\nu}}$.  (It can be verified that eq. (10) 
satisfies eq. (2).) The usual ``running coupling function'' $\overline{\lambda}(t)$ has the boundary condition 
$\overline{\lambda}(0) = \lambda_0$ with $\lambda_0$ also being equal to $\lambda(\eta = 0)$.  
It is now apparent that eqs. (7) and (10) are identical provided $\eta = 0$, and so we now have the boundary conditions for eq. (4)
\begin{equation}
T_{n0} = -\frac{1}{4}\,\delta_{n0}.
\end{equation}
Upon equating $L$ in eqs. (10) and (3) we see that $\frac{1}{\overline{\lambda}^2(t)} = \frac{-4}{\lambda_0^2} \left[\displaystyle{\sum_{n=0}^\infty} S_n(\lambda_0^2t)\lambda_0^{2n}\right]$, 
which is a novel expression for the running coupling in terms of $\beta(\lambda)$.  
Consequently the log-independent contributions to $L$ are fixed by the trace anomaly.

The effective potential $V$ for a massless scalar field with the classical potential $V_{C1} = \lambda\phi^4$ can be 
treated in an analogous fashion. We will now review how [12] the RG equation can be used to express the log-dependent part of $V$ in terms of the log-independent parts, 
and how these log-independent parts can be determined by considering an extra condition (which in this case is $V^\prime(\phi = v) = 0)$.  The expansion 
\begin{equation}
V = \sum_{n=0}^\infty \,\sum_{m=0}^n \lambda^{n+1} T_{n,m}L^m\phi^4\quad \left(L = \log \frac{\phi}{\mu}\right)
\end{equation}
when expressed as
\begin{equation}
V = \sum_{n=0}^\infty A_n(\lambda)L^n\phi^4
\end{equation}
(where $A_n = \displaystyle{\sum_{m=n}^\infty} T_{m,n}\lambda^{m+1}$) satisfies the RG equation
\begin{equation}
\left(\mu \frac{\partial}{\partial\mu} + \beta(\lambda) \frac{\partial}{\partial\lambda} + \gamma(\lambda)\phi \frac{\partial}{\partial\phi}\right)V = 0
\end{equation}
provided
\begin{equation}
A_{n+1}(\lambda) = \frac{1}{n+1}\left(\hat{\beta}\frac{\partial}{\partial\lambda} + 4\hat{\gamma}\right)A_n(\lambda)
\end{equation}
where $\hat{\beta} = \beta/(1-\gamma)$ and $\hat{\gamma} = \gamma/(1-\gamma)$.  If now 
\begin{equation}
\eta = \int_{\lambda_0}^{\lambda(\eta)} \,\frac{dx}{\hat{\beta}(x)}
\end{equation}
and
\begin{equation}
\hat{A}_n(\lambda) = A_n(\lambda)\exp \left(4 \int_{\lambda_0}^\lambda \frac{\hat{\gamma}(x)}{\hat{\beta}(x)} dx\right)
\end{equation}
then by eq. (15)
\begin{equation}
\hat{A}_{n+1}(\lambda(\eta)) = \frac{1}{n+1} \frac{d}{d\eta}\hat{A}_n (\lambda(\eta)) = \frac{1}{(n+1)!} \frac{d^{n+1}}{d\eta^{n+1}} \hat{A}_0(\lambda(\eta)).
\end{equation}
The sum of eq. (13) now leads to 
\begin{equation}
V = A_0(\lambda(\eta + L)) \exp \left(4 \int_{\lambda(\eta)}^{\lambda(\eta + L)} \frac{\gamma(x)}{\beta(x)} dx\right)\phi^4.
\end{equation}
As with eq. (7), eq. (19) shows that effective potential is determined by its log-independent contributions and the RG functions.

To fix these log-independent contributions to $V$, we need a second condition.  The trace of the energy-momentum tensor does not help us to do this.  
However, we can invoke the condition
\begin{equation}
\frac{dV(\phi)}{d\phi}\left|_{\phi=v} = 0\right.
\end{equation}
where $v$ is the vacuum expectation value of $V$.  If the renormalization scale parameter $\mu$ is chosen to be equal to $v$, then by eqs. (13) and (20)
\begin{equation}
[A_1(\lambda) + 4A_0(\lambda)]v^3 = 0.
\end{equation}
This equation has been derived for a particular value of $\mu$, but as $\lambda$ at this value of $\mu$ is not fixed, eq. (21) implies the functional relation
\begin{equation}
A_1(\lambda) = -4A_0(\lambda)
\end{equation}
provided $v \neq 0$.  Eq. (22) and eq. (15) with $n = 0$ together lead to
\begin{equation}
\left[\hat{\beta} \frac{d}{d\lambda} + 4(1 + \hat{\gamma})\right]A_0 = 0
\end{equation}
so that 
\begin{equation}
A_0(\lambda) = A_0(\lambda_0) \exp \left(-4 \int_{\lambda{_0}}^\lambda \frac{dx}{\beta(x)}\right),
\end{equation}
and hence eq. (19) becomes 
\begin{align}
V &= A_0(\lambda_0)\exp\left(-4 \int_{\lambda_0}^{\lambda(\eta + L)} \,\frac{dx}{\beta(x)}\right)
\exp \left(4 \int_{\lambda_{(\eta)}}^{\lambda(\eta + L)} \,\frac{\gamma(x)}{\beta(x)} dx\right) \phi^4\nonumber\\
&= A_0 (\lambda_0) \exp \left(-4 \int_{\lambda{_0}}^\lambda \frac{dx}{\beta(x)}\right)\mu^4 ,
\end{align}
upon using eq. (16).  
Consequently, $V$ is independent of $\phi$ provided $v \neq 0$; either there is no spontaneous symmetry breakdown or the potential is ``flat''.  (Of course, this flatness does 
not preclude spontaneous symmetry breaking.)

We thus see that the effective Lagrangian for a constant gauge field and the effective potential for a massless scalar field are completely determined by the RG functions when the RG equation is supplemented by a suitable extra condition. In the case of the effective action for an external electromagnetic field this extra 
condition is provided by the anomalous trace of the energy momentum tensor.  For the effective potential $V(\phi)$ in a massless $\phi^4$ model, it is the fact that $V^\prime(0)$ 
disappears when $\phi = v$ which determines $V(\phi)$ completely in terms of the RG functions.

\section{Acknowledgments}
The author would like to thank F.T. Brandt, F. Chishtie, T. Hanif, J. Jia, T.N. Sherry 
and C. Schubert for discussions.  R. Macleod had a helpful suggestion.

\end{document}